\documentclass[twocolumn]{aastex63}
\usepackage[utf8]{inputenc}
\usepackage{amsmath,amsfonts,amssymb}
\usepackage{graphicx,subfigure,multirow}
\usepackage{epsfig,color}
\usepackage{enumitem}
\usepackage{natbib}
\usepackage{hyperref}
\usepackage{verbatim}

\begin{document}
\title{Impact of PREX-II and combined Radio/NICER/XMM-Newton's mass-radius measurement of PSRJ0740+6620 on the dense matter equation of state}

\author[0000-0003-2131-1476]{Bhaskar Biswas}
\affiliation{Inter-University Centre for Astronomy and Astrophysics, Post Bag 4, Ganeshkhind, Pune 411 007, India}

\begin{abstract}
In this paper, we discuss the impact of the following laboratory experiments and astrophysical observation of neutron stars (NSs) on its equation of state (EoS): (a) The new measurement of neutron skin thickness of $\rm ^{208} \! Pb$, $R_{\rm skin}^{208} = 0.29 \pm 0.07$ fm by the PREX-II experiment. (b) The mass measurement of PSR J0740+6620 has been slightly revised down by including additional $\sim 1.5$ years of pulsar timing data. As well as the radius measurement of PSR J0740+6620 by joint NICER/XMM-Newton collaboration which has a similar size to PSR J0030+0451. We combine these information using Bayesian statistics along with the previous LIGO/Virgo and NICER observations of NS using a hybrid nuclear+piecewise polytrope (PP) EoS parameterization. Our findings are as follows: (a). Adding PREX-II result yields the value of empirical parameter $L = 69^{+21}_{-19}$ MeV, $R_{\rm skin}^{208} = 0.20_{-0.05}^{+0.05}$ fm, and radius of a $1.4 M_{\odot}$ ($R_{1.4}) = 12.75_{-0.54}^{+ 0.42}$ km at $1 \sigma$ confidence interval (CI). We find these inferred values are mostly dominated by the combined astrophysical observations as the measurement uncertainty in $R_{\rm skin}^{208}$ by PREX-II is much broader. Also, a better measurement of $R_{\rm skin}^{208}$ might have a little effect on the radius of low mass NSs, but for the high masses there will be almost no effect. (b) After adding The revised mass and radius measurement of PSR J0740+6620, we find the inferred radius of NSs is slightly pushed towards the larger values and the uncertainty on the radius of a $2.08 M_{\odot}$ NS is moderately improved.

\end{abstract}

 \section{Introduction} Throughout the last few decades, understanding dense matter EoS has been one of the most key challenges to multiple physics and astrophysics communities. Observation of macroscopic properties of a NS such as mass, radius, tidal deformability, moment of inertia could provide us fascinating information about the dense matter EoS.  Thanks to the LIGO/Virgo collaboration~\citep{advanced-ligo,advanced-virgo}, we have now entered into multi-messenger era, in which we have already observed gravitational wave (GW) signals from multiple likely binary neutron star (BNS) merger system~\citep{TheLIGOScientific:2017qsa,Abbott:2018exr,Abbott:2020uma}. In 2019, NICER collaboration~\citep{2016SPIE.9905E..1HG} for the first time has reported a very accurate measurement of mass and radius of PSR J0030+0451~\citep{Riley:2019yda,Miller:2019cac} by observing X-ray emission from several hot spots of NS surface. Additionally, very accurate mass measurement of NS by the radio observations, particularly the heaviest one~\citep{Cromartie:2019kug} also helps us a lot to constrain the high-density EoS. In a recent work~\citep{Biswas_arXiv_2008.01582B} (also see other works~\citep{Raaijmakers:2019dks,Landry_2020PhRvD.101l3007L,Jiang:2019rcw,Traversi:2020aaa,Al-Mamun:2020vzu,Dietrich:2020efo} using different EoS parameterization), using Bayesian statistics we have already combined the aforementioned observations based on a hybrid nuclear+PP EoS parameterization and placed a stringent constraint on the NS EoS. This hybrid parameterization is constructed by combining two widely used EoS models: near the saturation density ($\rho_0$) nuclear empirical parameterization~\citep{Piekarewicz:2008nh,Margueron:2017eqc} is used based on a parabolic expansion of energy per nucleon and at higher densities a nuclear physics agnostic PP parameterization~\citep{Read:2008iy}, as the high-density EoS could not be probed by the current nuclear physics understanding. This hybrid parameterization is also used recently to investigate the nature of the ``mass-gap'' object in GW190814~\citep{Biswas:2020xna}.

In recent times, laboratory experiments such as PREX-II have shown us a promise to put further constraint~\citep{Reed:2021nqk,Essick:2021kjb} on the NS EoS. They have reported~\citep{Adhikari:2021phr} the value of neutron skin thickness of $\rm ^{208} \! Pb$ to be, $R_{\rm skin}^{208} = 0.29 \pm 0.07$ fm (mean and $1 \sigma$ standard deviation). Such a measurement can give us crucial information about the nuclear EoS around the saturation density. In particular, our hybrid nuclear+PP model directly allows us to include the result obtained from the PREX-II experiment as the empirical parameter like $L$ shows a strong correlation with $R_{\rm skin}^{208}$~\citep{Vinas:2013hua,Reinhard:2016mdi}. So the aim of this paper is to improve our knowledge on dense matter EoS using hierarchical Bayesian statistics by including this newly obtained result from PREX-II experiments (combining with other aforementioned observations) under the hybrid nuclear+PP EoS parameterization.

Finally, this is to note that very recently the mass measurement of PSR J0740+6620 has been revised down by including additional $\sim 1.5$ years of pulsar timing data~\citep{2021arXiv210400880F}. Interestingly, NICER collaboration has also been taking data of this object using the X-ray pulse profile modelling and they are able to measure the radius of this object as well. Additionally, to improve the total flux measurement of the star, they also include X-ray multi mirror (XMM)-Newton telescope~\citep{Turner:2000jy,struder:2001} data which have a far smaller rate of background counts than NICER. Two independent analysis using this joint NICER/XMM-Newton data have estimated the radius to be $12.39^{+1.30}_{-0.98}$ km~\citep{Riley:2021pdl} and $13.71^{+2.61}_{-1.50}$ km~\citep{Miller:2021qha}. In this paper, we use the mass-radius estimate of PSR J0740+6620 both from~\citep{Riley:2021pdl} and~\citep{Miller:2021qha}, and study its impact on the dense matter EoS.    


 \section{Hybrid nuclear+PP EoS inference methodology} Hybrid nuclear+PP is an EoS model which connects a nuclear physics informed EoS parameterization in the vicinity of $\rho_0$ with a high-density nuclear physics agnostic PP EoS parameterization. Since the crust has a very minimal impact on NS macroscopic properties~\citep{Biswas:2019ifs,Gamba:2019kwu}, standard BPS EoS~\citep{1971ApJ...170..299B} is used in this model for the low-density regime and joined with the core EoS in a thermodynamically consistent fashion~\citep{Xie:2019sqb}. EoS around $\rho_0$ is well described via parabolic expansion of energy per nucleon and can be divided into two parts:
\begin{equation}
    e(\rho,\delta) \approx  e_0(\rho) +  e_{\rm sym}(\rho)\delta^2,
\end{equation}
the term $e_0(\rho)$ corresponds to the energy of symmetric nuclear matter for which the number of neutrons is equal to the number of protons, $e_{\rm sym}(\rho)$ is the energy of the asymmetric nuclear matter or widely named as symmetry energy, and $\delta=(\rho_n-\rho_p)/\rho$ is known as symmetry parameter where $\rho_n$, $\rho_p$, and $\rho$ are respectively the number density of neutron, proton, and the total number density. Around $\rho_0$, this two energy can be further expanded in a Taylor series: 
\begin{eqnarray}
 e_0(\rho) &=&  e_0(\rho_0) + \frac{ K_0}{2}\chi^2 \label{eq:e0} +\,...,\\
e_{\rm sym}(\rho) &=&  e_{\rm sym}(\rho_0) + L\chi + \frac{ K_{\rm sym}}{2}\chi^2 
 ..., \label{eq:esym}
\end{eqnarray}
where $\chi \equiv (\rho-\rho_0)/3\rho_0$ quantifies deviation from saturation density which must be always much smaller than unity. In this work, we truncate the Taylor expansion up to the second-order in $\chi$. As the lowest order parameters are experimentally well determined, we fix them in this analysis, such as $e_0(\rho_0) = -15.9$ MeV, and $\rho_0 =0.16 \rm{fm^{-3}}$. The symmetry energy at $\rho_0$, curvature ($K_0$) of symmetric matter, slope ($L$), and the curvature ($K_{\rm sym}$) of symmetry energy are the four free parameters in the model. Above $1.25 \rho_0$ density the nuclear empirical parameterization is no longer taken to be reliable (for details, see ~\citep{Biswas_arXiv_2008.01582B}) and joined with 3-segment PP EoS parameterization ($\Gamma_1$, $\Gamma_2$, and $\Gamma_3$ be the three polytropic indices). The prior ranges for all the free parameters in this hybrid nuclear+PP model are shown in~\ref{tab:prior}. One notable difference from Ref.~\citep{Biswas_arXiv_2008.01582B} is now we have taken broader prior for the high-density parameters to avoid such situation where priors could rail against the posteriors. However the priors for the nuclear-physics informed parameters such as $K_0$, $e_{\rm sym}$, and $L$ remains to be same as the previous work.

\begin{table}[ht]
    \centering
    \begin{tabular}{c|c}
         \hline
         Parameter& Prior  \\
         \hline
         $  K_0$ (MeV) & $\mathcal{N}(240,30)$ \\
         $ e_{\rm sym}$ (MeV) & $\mathcal{N}(31.7,3.2)$ \\
         $L$ (MeV) & $\mathcal{N}(58.7,28.1)$ \\
         $ K_{\rm sym}$ (MeV) & uniform(-1000,500) \\
         $ \Gamma_1$  & uniform(0.2,8)\\
         $ \Gamma_2$  & uniform(0.2,8)\\
         $ \Gamma_3$  & uniform(0.2,8)\\
         \hline
    \end{tabular}
    \caption{Prior ranges of various EoS parameters. A Gaussian prior on $K_0$, $e_{\rm sym}$, and $L$ is considered here indiacting their mean and $1 \sigma$ CI. For the other four parameters, wide uniform priors are assumed.}
    \label{tab:prior}
\end{table}

\begin{figure*}[ht!]
\begin{tabular}{cc}
\includegraphics[width=0.45\textwidth]{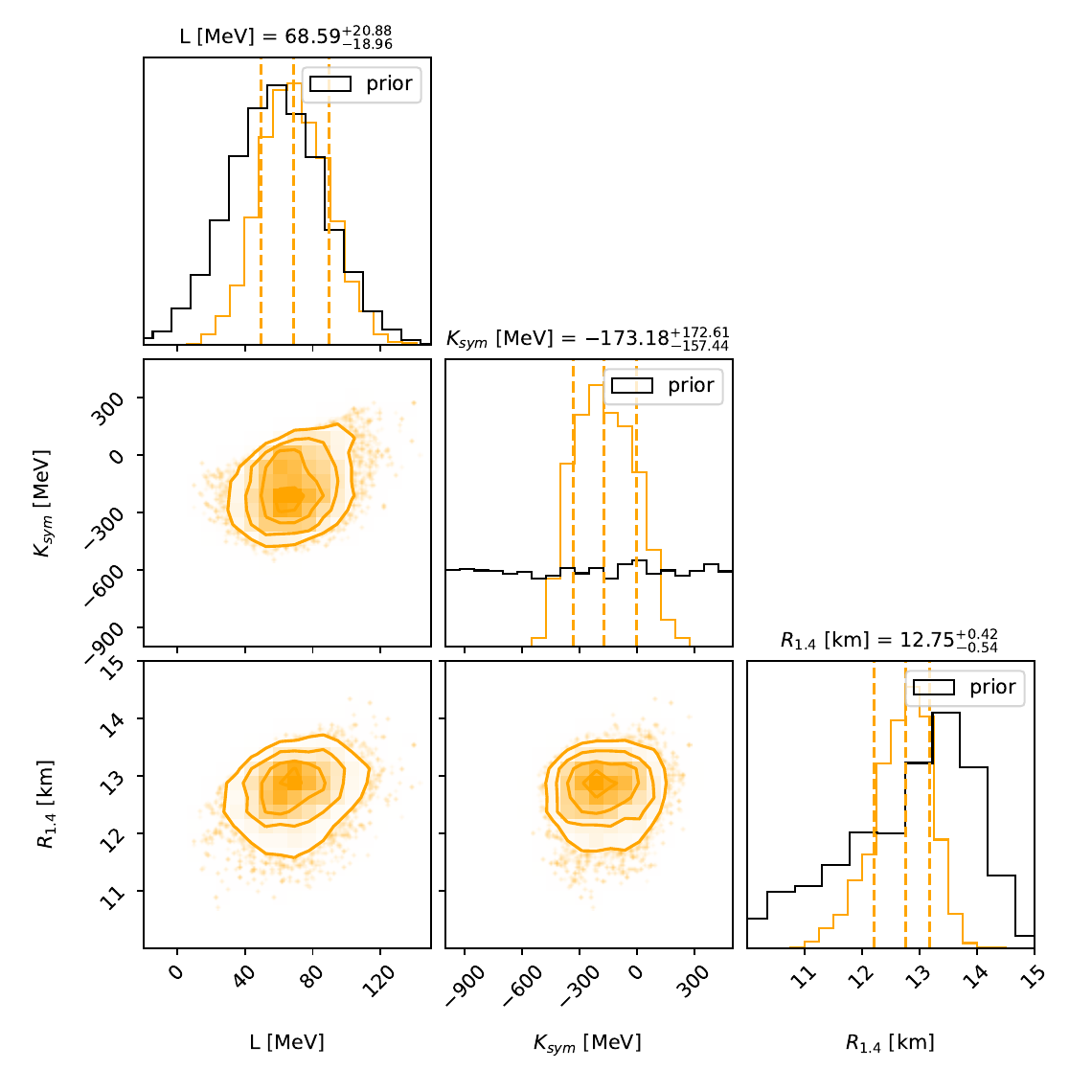}&
\includegraphics[width=0.45\textwidth]{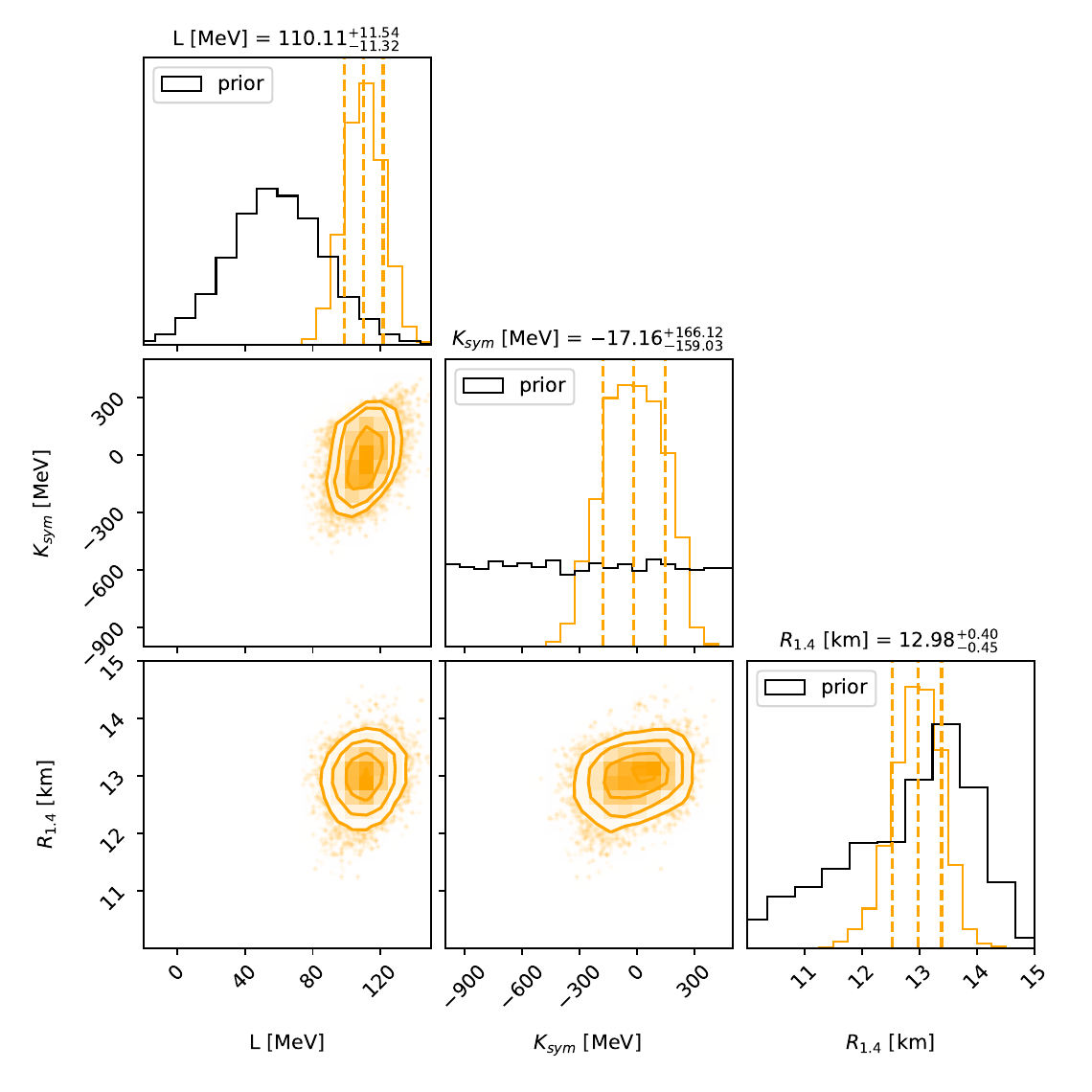}\\

\end{tabular}
\caption{(a). In the left panel, the posterior distribution of empirical parameters $L$ and $K_{\rm sym}$, and their correlation with $R_{1.4}$ are shown after adding PREX-II result with astrophysical observations. In the marginalized one-dimensional plot corresponding prior, median, and $1 \sigma$ CI are also shown. (b.) The right panel is the same as the left but a hypothetical measurement of neutron skin thickness of $\rm ^{208} \! Pb$ $R_{\rm skin}^{208} = 0.29_{-0.02}^{+0.02}$ is added with with astrophysical observations. }
\label{fig:params-dist}
\end{figure*}

Then the posterior of the EoS parameters (denoted as $\theta$) are computed through nested sampling algorithm implemented in~{\tt Pymultinest}~\citep{Buchner:2014nha}:
\begin{equation}
    P(\theta | {d}) \propto P(\theta) \Pi_i P ({d_i} | \theta) \,,
    \label{bayes theorem}
\end{equation}
where $d = (d_1, d_2,...)$ is the set of data from different 
types of experiments and observations, $P ({d_i} | \theta)$ are corresponding likelihood distribution, and $P(\theta)$ are the priors on the EoS parameters $\theta$. For astrophysical observations, likelihood distributions are modelled in the following fashion: (a) Old mass measurement of PSR J0740+6620~\citep{Cromartie:2019kug} is modelled with a Gaussian likelihood of $2.14 M_{\odot}$ mean and $0.1 M_{\odot}$ $1 \sigma$ standard deviation. (b) Mass and tidal deformability measurement from GW170817~\citep{TheLIGOScientific:2017qsa} and GW190425~\citep{Abbott:2020uma} are modelled with Gaussian kernel density estimator (KDE). (c) Similarly mass and radius measurement of of PSR J0030+0451~\citep{Riley:2019yda,Miller:2019cac} and PSR J0740+6620~\citep{Riley:2021pdl,Miller:2021qha} are also modelled with Gaussian KDE. For the PREX-II experiment, the likelihood function is taken to be a Gaussian distribution of skin thickness with $0.29 \pm 0.07$ fm (mean and $1 \sigma$ standard deviation). Similar to Ref.~\citep{Essick:2021kjb} for the likelihood computation of PREX-II, we use the following universal relation obtained from Ref.~\citep{Vinas:2013hua} between $r_{\rm skin}$ and empirical parameter $L$: $R_{\rm skin}^{208} {\rm [fm]} = 0.101 + 0.00147 \times L [\rm MeV]$.

\begin{figure}[ht]
    \centering
    \includegraphics[width=.45\textwidth]{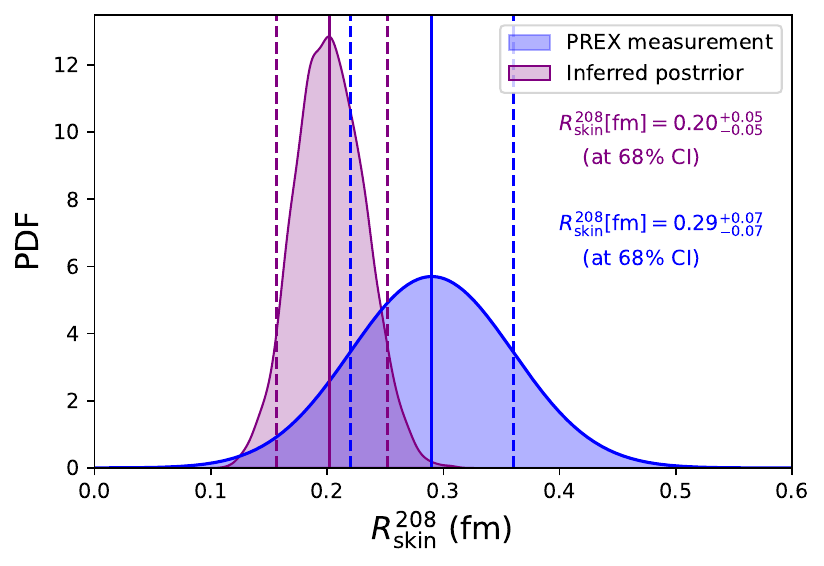}
    \caption{In the purple shade, the inferred posterior distribution of $R_{\rm skin}^{208}$ is shown using joint astrophysical+PREX-II data. In blue shade, the PREX-II measured distribution is shown. For both of the distributions the corresponding median and $1 \sigma$ CI are also indicated using solid and dotted lines in the same colour respectively.  }
    \label{fig:Rskin-dist}
\end{figure}

\begin{figure*}[ht]
    \centering
    \includegraphics[width=\textwidth]{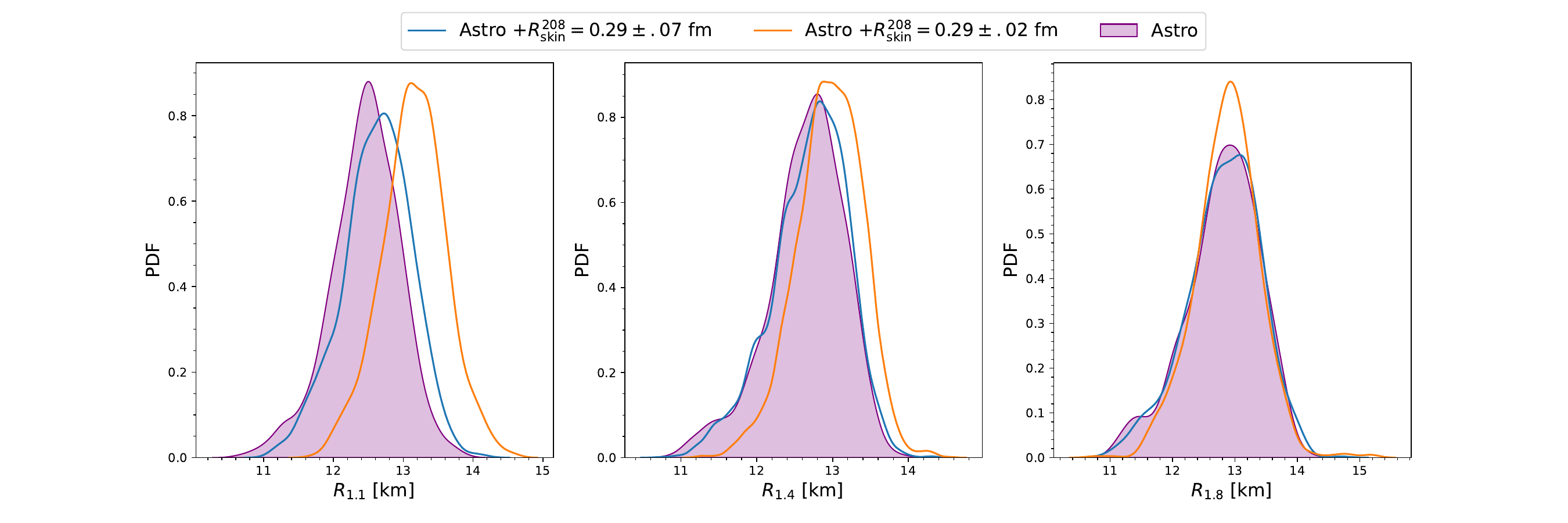}
    \caption{The posterior distributions of radius are shown for three different NS masses: 1.1 (left), 1.4 (moddle), and 1.8 $M_{\odot}$ (right).   }
    \label{fig:Rskin-radius}
\end{figure*}


\begin{figure*}[ht]
    \centering
    \includegraphics[width=\textwidth]{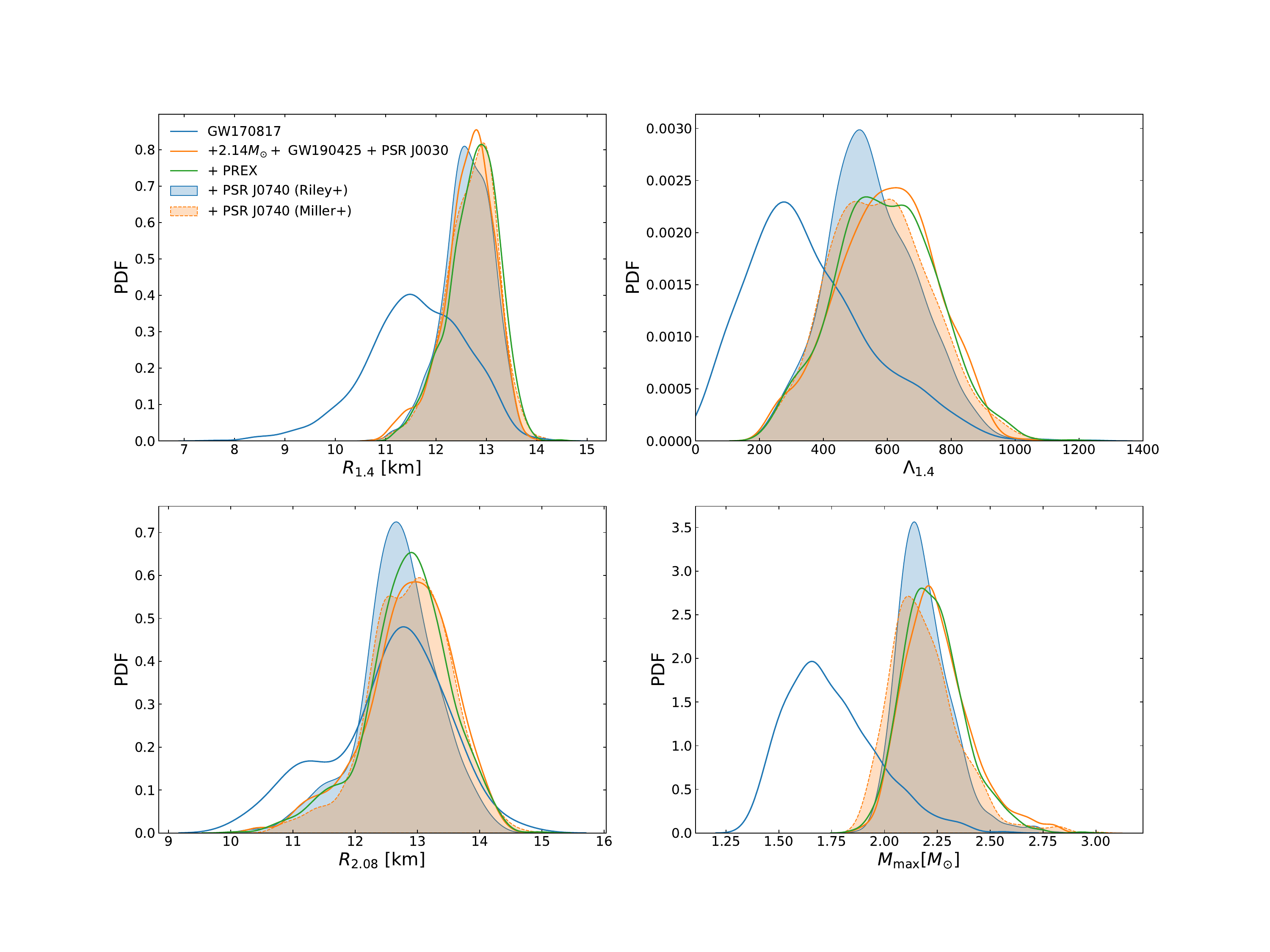}
    \caption{Inferred posterior distribution of various macroscopic properties such as $R_{1.4}$, $\Lambda_{1.4}$, $R_{2.08}$, and $M_{\rm max}$ are shown adding successive observations.}
    \label{fig:macrro-pop}
\end{figure*}

\begin{figure*}[ht!]
    \centering
    \includegraphics[width=\textwidth]{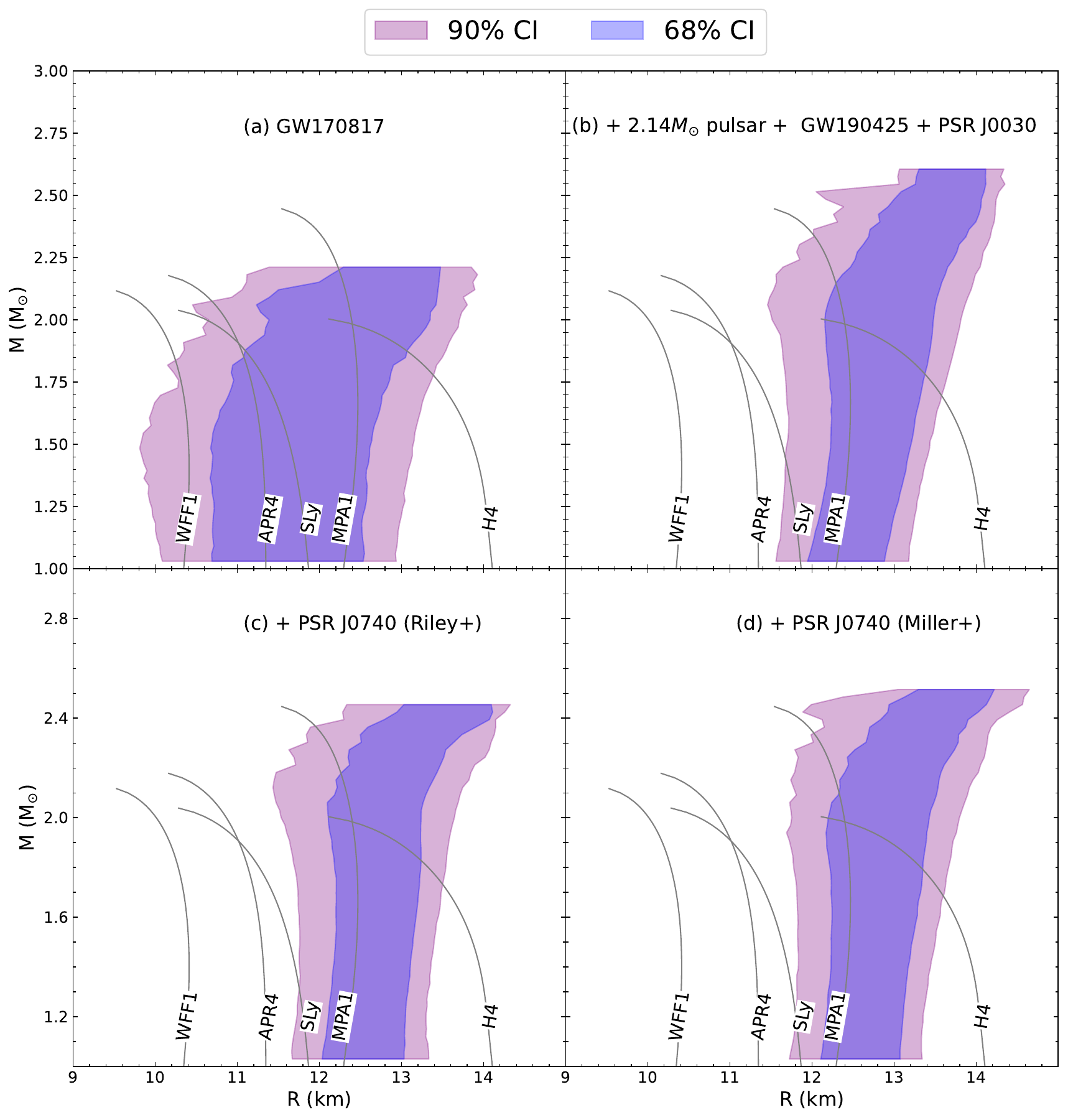}
\caption{$68 \%$ and $90 \%$ CI of mass-radius are shown: In panel a, constraints coming from  GW170817 observation is shown. In panel b, mass measurement ($2.14 \pm 0.1 M_{\odot}$) of PSR J0740+6620, GW190425, and PSR J0030+0451 are added. In panel c and d mass-radius measurement of PSR J0740+6620 from~\citet{Riley:2021pdl} and~\citet{Miller:2021qha} are added respectively.} Some standard EoS curves like APR \citep{1998PhRvC..58.1804A}, SLy \citep{Douchin:2001sv}, WFF1 \citep{1988PhRvC..38.1010W}, MPA1 \citep{1987PhLB..199..469M}, and H4 \citep{2006PhRvD..73b4021L} are also overlaid.
    \label{fig:MR-post-comparison}
\end{figure*}

\section{Results} In the left panel of Fig.~\ref{fig:params-dist}, the posterior distribution of the empirical parameters $L$ and $K_{\rm sym}$, and their correlation with $R_{1.4}$ are shown after combining astrophysical observations (old mass measurement of PSR J0740+6620, GWs, and PSR J0030+0451) and PREX-II experiment results. In the marginalized one-dimensional plot corresponding prior, median, and $1 \sigma$ CI are also given. Posterior distribution of $K_0$ and $e_{\rm sym}$ are not shown here as the current data is unable to provide any significant information about them. However, the constraint on $L$ and $K_{\rm sym}$ is improved. Before adding PREX-II results, the bound on $L$ was $54^{+21}_{-20}$ MeV at $1 \sigma$ CI and now it becomes to be $69^{+21}_{-19}$ MeV. Constraint on $K_{\rm sym}$ is not as strong as $L$, but data suggests the negative value of $K_{\rm sym}$. We find a very weak correlation between $R_{1.4}$ and $L$ (of about $\sim 0.27$), and therefore the PREX-II data has very marginal effect on $R_{1.4}$. Ref.~\citep{Reed:2021nqk} has extracted a much larger ($106 \pm 37$ MeV) value of $L$ compared to us using the PREX-II result and they also obtain a strong correlation between $R_{1.4}$ and $L$. Because of this strong correlation, they find the radius and tidal deformability of the NS have to be large which reveals tension with the GW170817 result. The reasons for this discrepancy are as follows: (a) To deduce the properties of NS one must combine all the available information using hierarchical Bayesian statistics. Ref.~\citep{Reed:2021nqk} deduce a larger radius for NS based on PREX-II result alone using a correlation-based study and later they find the result is in tension with GW170817. (b) Ref.~\citep{Reed:2021nqk} also use a handful number of EoSs that can introduce model dependence as mentioned by~\citet{Essick:2021kjb}. On the other hand, our result is robust as we sample the posterior distribution of EoS parameters directly using the data using a nested sampling algorithm.


In Fig.~\ref{fig:Rskin-dist}, the inferred posterior distribution of $R_{\rm skin}^{208}$ (in purple shade) which is obtained after combining astro+PREX-II data is compared with the PREX-II measured distribution (in blue shade). Though the uncertainties of these two distributions overlap with each other within the $1 \sigma$ CI, we find the inferred value of $R_{\rm skin}^{208} = 0.20_{-0.05}^{+0.05}$ fm is relatively smaller than PREX-II measured value. This smaller inferred value of $R_{\rm skin}^{208}$ is consistent with Ref.~\citep{Xu:2020fdc,Essick:2021kjb,Tang:2021snt,Li:2021thg} using different assumption of EoS parameterization. This suggests that there is a mild tension between astrophysical observations and PREX-II data. Currently astrophysical observations dominate over the PREX-II measurement. The uncertainty in the measurement of $R_{\rm skin}^{208}$ by PREX-II is still much broader. If the high value of $R_{\rm skin}^{208}$ persists with lesser uncertainty then this tension will be revealed. To support this statement we consider a hypothetical measurement of $R_{\rm skin}^{208} = 0.29 \pm 0.02$ fm (median and $ 1\sigma$ CI) and  combined with the existing astrophysical observations. With this lesser uncertainty in the measured $R_{\rm skin}^{208}$, we find the inferred values of $L = 110_{-10}^{+11}$ MeV and $R_{\rm skin}^{208} = 0.26_{-0.02}^{+0.03}$ fm. These inferred values would pose a challenge to the current theoretical understanding about the nuclear matter near the saturation densities. In the right panel of Fig.~\ref{fig:params-dist}, the posterior distributions of $L$ and $K_{\rm sym}$, and their correlations with $R_{1.4}$ are shown for this hypothetical case. This time we find a slightly higher value ($\sim .2$ km) of $R_{1.4}$ compare to the predicted $R_{1.4}$ by combined astro+PREX-II data. We further check how the skin thickness measurement affects the radius of different NS masses. In Fig.~\ref{fig:Rskin-radius} the distribution of radius of 1.1,1.4, and 1.8 $M_{\odot}$ NSs are shown using three (astro, astro+PREX-II, and astro+hypothetical $R_{\rm skin}^{208}$) different types of data. We see virtually no change in $R_{1.4}$ and $R_{1.8}$ when PREX-II measurement is added with other astrophysical observations. $R_{1.1}$ is slightly increased by the addition of PREX-II data. This increment gets moderately higher when the hypothetical measurement of $R_{\rm skin}^{208}$ is added. For $R_{1.4}$, we see only a slight increment even after adding the hypothetical measurement of $R_{\rm skin}^{208}$ and almost no change for $R_{1.8}$. Therefore, a better measurement of might have a small effect on the radius of low masses NS, but for the high masses NSs there will be almost no effect. A similar conclusion has also been achieved recently in~\citep{Essick:2021ezp} using their nonparametric equation of state representation based on Gaussian processes.

\begin{table*}[ht]
\centering
\begin{tabular}{|p{2cm}|p{2.5cm}|p{2.5cm}|p{2.5cm}|p{2.5cm}|} 
\hline

 Quantity&  GW170817 & $+2.14 M_{\odot}$ pulsar \newline  GW190425$+$\newline PSR J0030 & $+$PSR J0740 \newline (Riley+) & $+$PSR J0740 \newline (Miller+)\\

\hline 
\vspace{1ex}

 $R_{1.4} [\rm km]$ &$11.61_{-1.72}^{+1.45}  $ & $12.49_{-0.90}^{+0.69}$ &  
  $12.64_{-0.88}^{+0.71}$ & $12.75_{-0.92}^{+0.68}$\\

\vspace{1ex}
 $R_{2.08} [\rm km]$ &$12.66_{-1.73}^{+1.08}  $& $12.91_{-1.43}^{+0.97}$ & 
   $12.70_{-1.25}^{+1.01}$ &$12.92_{-1.11}^{+0.97}$\\

\vspace{1ex}
 $\Lambda_{1.4}$ & $322_{-229}^{+391}$ & $602_{-267}^{+241}$ & 
   $538_{-211}^{+249}$ &$575_{-232}^{+262}  $\\

\vspace{1ex}
 $M_{\rm max} (M_{\odot})$ & $1.71_{-0.25}^{+0.42}$ & $2.22_{-0.19}^{+0.31}$ & 
   $2.17_{-0.15}^{+0.24}$ &$2.16_{-0.19}^{+0.30}  $\\

\hline 
\end{tabular}
\caption{Median and $90 \%$ CI of $R_{1.4}$, $R_{2.08}$, $\Lambda_{1.4}$, and $M_{\rm max} (M_{\odot})$ are quoted here after adding successive observations.   
}
\label{tab-radius}*
\end{table*}

Finally we discuss the impact of the radius measurement of PSR J0740+6620 by joint NICER/XMM-Newton data. Since the uncertainties in the radius measurement of PSR J0740+6620 by~\citet{Miller:2021qha} is larger than the~\citet{Riley:2021pdl} due to a conservative treatment of the calibration error, we analyse both data separately and compare the results.
In Fig.~\ref{fig:MR-post-comparison}, $68 \%$ (in blue shade) and $90 \%$ CI (in purple shade) of mass-radius posterior is shown adding successive observations. Also some reference EoSs are plotted in solid black lines. In the upper left panel, constraints coming from GW170817 observation is shown. Also the corresponding various macroscopic properties such the posterior distribution of $R_{1.4}$, $R_{2.08}$, $\Lambda_{1.4}$, and $M_{\rm max} (M_{\odot})$ are shown Fig.~\ref{fig:macrro-pop} and their median and $90\%$ CI are quoted in Table~\ref{tab-radius}. Interestingly, the LIGO-Virgo published GW170817 posterior of $\widetilde{\Lambda}$ has a bi-modality: The primary mode peaks at $\widetilde{\Lambda} \sim$ 200 and the secondary one peaks at $\widetilde{\Lambda} \sim$ 600. We also see the inferred posterior of $\Lambda_{1.4}$ peaks around $\sim 200$ using the GW170817 observation alone. After adding the mass measurement of PSR J0740+6620 ($2.14 \pm 0.1 M_{\odot}$), GW190425, and PSR J0030+0451, we find the combined data no longer favors the primary mode of GW170817 but it only favors the secondary mode (see fig.~\ref{fig:Rskin-radius}). In the panel c and d of fig.~\ref{fig:MR-post-comparison} we show the resulting mass-radius posterior due to the addition of PSR J07400+0620 using the data from~\citet{Riley:2021pdl} and~\citet{Miller:2021qha} respectively. We find for the low masses NSs, both data result into similar bound of radius. There is only $\sim 0.07 (0.36)$ km differences in $R_{1.4} (R_{2.08})$ towards the stiff EoSs when we add data from~\citet{Miller:2021qha} instead of~\citep{Riley:2021pdl}. Also it can be noticed the estimated  $M_{\rm max}$ is slightly lower after adding PSR J07400+0620 due to its revised mass estimate.

A direct comparison between our results and other studies~\citep{Raaijmakers:2021uju,Miller:2021qha,Pang:2021jta,legred2021impact} can be done due to the different model assumptions and choice of different combinations of data sets. \citet{Raaijmakers:2021uju} reports $R_{1.4}= 12.33^{+0.76}_{-0.81}$ km and
$M_{\rm max}=  2.23^{+0.14}_{-0.23}M_{\odot} $ at the 95\% CI based on piecewise-polytrope parameterization which is informed by chiral-effective theory calculations at lower densities. ~\citet{legred2021impact} reports $R_{1.4}=12.54^{+1.01}_{-1.06}(12.32^{+1.02}_{-1.23})$ km and $M_{\rm max}= 2.24^{+0.34}_{-0.24}(2.22^{+0.30}_{-0.21})M_{\odot} $ at $90\%$ CI employing nonparametric EoS model based on Gaussian process using the data from~\citet{Miller:2021qha}(~\citet{Riley:2021pdl}).~\citep{Miller:2021qha} also use Gaussian process model in their analysis and their results are close to~\citep{legred2021impact}. We find our results are in good agreement with them but with a slightly higher median value of $R_{1.4}$ of about maximum $\sim 0.3$ km. This is due to the additional constraints coming from the heavy-ion collision results which are used to inform our nuclear empirical parameterization. ~\citet{Pang:2021jta} finds $R_{1.4}=12.03^{+0.77}_{-0.87}$ km and $M_{\rm max}=2.18^{+0.15}_{-0.15} M_{\odot}$ at the 90\% CI using chiral-effective theory-informed analysis, which are much smaller compared to our results. This is mainly due to the fact that ~\citep{Pang:2021jta} use a discrete sampling method consisting of 5000 EoSs and we use a nested sampling algorithm. While the discrete sampling algorithm is more time- and cost-efficient but it provides less statistical certainty than the nested or Monte Carlo sampling algorithm. Also, they include the electromagnetic counterpart of GW170817 which we do take into account here due to its various systematic uncertainties.

\section{Conclusion} In summary, we have investigated the impact of the recent PREX-II experimental results, the revised mass measurement of PSR J0740+6620, and as well as its radius measurement on the dense matter EoS based on a hybrid nuclear+PP EoS parameterization. The PREX-II data combining with astrophysical observations predict a slightly larger value of $L$ compared to when we only use astrophysical observations. However, the value of $L$ is in good agreement with other experimental determinations and theoretical expectation~\citep{Margueron:2017eqc}. We find a very weak correlation between $L$ and $R_{1.4}$ which does not change the radius much. We also argue that the dominant contribution on the inferred EoS posterior comes from the combined astrophysical observations as the measurement uncertainty in $R_{\rm skin}^{208}$ by PREX-II is still much broader. It is also shown that a better measurement of $R_{\rm skin}^{208}$ could have a little effect on the radius of low mass NSs, but the effect on the radius of high mass NSs will be almost negligible. Finally, we discuss the effect of the revised mass and radius measurement of PSR J0740+6620 using the data from both~\citep{Riley:2021pdl} and~\citep{Miller:2021qha}. Inferred radius using both data are broadly consistent with each other with a maximum of $\sim 0.36$ km differences in $R_{2.08}$ towards the stiff EoSs using the data from~\citep{Miller:2021qha}. The estimated $M_{\rm max}$ also gets slightly lower after adding PSR J0740+6620 mainly due to its revised mass measurement.

\section*{Acknowledgments} I thank the anonymous referee for the useful suggestions which has helped to improve this manuscript. I am indebted to Sukanta Bose, Rana Nandi, and Prasanta Char for useful discussions regarding this work. I gratefully acknowledge the use of high performance super-computing cluster Pegasus at IUCAA for this work. This material is based upon work supported by NSF’s LIGO Laboratory which is a major facility fully funded by the National Science Foundation.
\bibliography{mybiblio}
\end{document}